\documentclass[12pt,a4paper]{article}

\usepackage[utf8x]{inputenc}
\usepackage[english]{babel}
\usepackage{slashed}

\usepackage{xcolor}
\definecolor{dark}{rgb}{0.10,0.2,0.3}
\definecolor{magenta}{rgb}{0.7,0.1,0.3}
\definecolor{purpure}{rgb}{0.5,0.15,0.3}

\usepackage{graphicx}
\usepackage{dcolumn}
\usepackage{bm}
\usepackage{epstopdf}
\usepackage{mathrsfs}
\usepackage{amssymb,amsfonts,latexsym}
\usepackage{amsmath,bbold}
\allowdisplaybreaks 

\usepackage[breaklinks]{hyperref}
\hypersetup{colorlinks,
citecolor=blue,
filecolor=blue,
linkcolor=magenta,
urlcolor=purpure,hyperfootnotes=true,pdftex}

\usepackage{breakurl}

\long\def\comment#1{ }

\newcommand{\rmd}{{\rm d}}

\makeatletter
  \def\my@tag@font{\normalsize}
  \def\maketag@@@#1{\hbox{\m@th\normalfont\my@tag@font#1}}
  \let\amsmath@eqref\eqref
  \renewcommand\eqref[1]{{\let\my@tag@font\relax\amsmath@eqref{#1}}}
\makeatother

\title{\bf The next-to-leading order vertex for a forward jet plus a rapidity gap at high energies} 
\author{M. Hentschinski$^{1}$, J. D. Madrigal Mart{\'\i}nez$^2$,\\
 B. Murdaca$^3$, A. Sabio Vera$^{4,5}$\\ \\
{\small $^1$ Department of Physics, Brookhaven National Laboratory,}\\
{\small Upton, NY 11973, USA.}\\
{\small $^2$ Institut de Physique Th{\' e}orique, CEA Saclay,}\\ 
{\small F-91191 Gif-sur-Yvette, France.}\\
{\small $^3$ Dipartimento di Fisica, Universit{\`a} della Calabria \&}\\
{\small Istituto Nazionale di Fisica Nucleare, Gruppo Collegato di Cosenza,}\\
{\small I-87036 Arcavacata di Rende, Cosenza, Italy.}\\
{\small $^4$ Instituto de F{\' \i}sica Te{\' o}rica UAM/CSIC, Nicol{\'a}s Cabrera 15}\\ 
{\small \& U. Aut{\' o}noma de Madrid, E-28049 Madrid, Spain}\\
{\small $^5$ CERN,  Geneva, Switzerland}
}

\begin{document} 

\maketitle

\begin{center}
{\bf \large Abstract}
\end{center}

We present the results for the calculation of the forward jet vertex associated to a 
rapidity gap (coupling of a hard pomeron to the jet) in the Balitsky-Fadin-Kuraev-Lipatov (BFKL) formalism at 
next-to-leading order (NLO). We handle the real emission contributions making use of the high energy effective action proposed by Lipatov, valid for multi-Regge and 
quasi-multi-Regge kinematics. This result is important since it allows, together with 
the NLO non-forward gluon Green function, to perform NLO studies of jet production in diffractive events (Mueller-Tang dijets, as a well-known example).

\newpage

\section{Introduction}

In recent years we have developed useful techniques \cite{ours} to work with the high energy effective action proposed by Lev Lipatov \cite{effective} to calculate scattering amplitudes relevant in the multi-Regge limit. This action is based on the separation of the emitted particles into clusters widely apart in rapidity. These clusters are connected to each other by reggeized gluon propagators which act as non-Sudakov form factors generating regions in rapidity with no emissions. The value of the effective action is 
to account for the interactions of these reggeized gluons with usual quarks and 
gluons inside the emission clusters. The strong ordering in rapidity among the clusters allows for the efficient resummation of powers of rapidity, or large logarithms in the center-of-mass energy, which are present in the scattering 
amplitudes. Of course, the coefficients of these logarithms are correctly calculated when compared to exact evaluations of the scattering amplitudes, both in elastic and complicated inelastic cases. 

There are at least two technical details which we need to treat with care in this 
program. One of them is to avoid double counting when considering that some 
emissions are inside the cluster or in a neighboring one. For this we have introduced a subtraction and regularization method. The second complication is at loop level since the effective action is formulated in terms of non-local operators which introduce new ultraviolet divergencies which are not related to short distance physics but have a kinematical origin. In this case we introduce a cut-off in the loop integrations which 
we have proven to be related to powers of total rapidity. This cut-off can be interpreted as a small deformation of the light cone. This is a delicate procedure, 
specially when one needs to perform two or higher loop calculations since the 
integrals we encounter are not of the standard types investigated in the literature. 

Our proposed methods have been tested in well-known quantities. At two loop level  we have successfully reproduced the gluon Regge trajectory at NLO, first the simpler quark contributions \cite{trajectory1} and then the more complicated gluon ones \cite{trajectory2}. At one loop level with inelastic amplitudes we have calculated the jet vertex for the production 
of  a jet in the forward direction coupled to a reggeized gluon \cite{forward}. This configuration 
has minijet radiation associated to the forward jet and it is used in the calculation of the so-called Mueller-Navelet cross sections \cite{mn} which are playing an important role in the application of the BFKL formalism \cite{BFKL} to phenomenology at the Large Hadron 
Collider (LHC). In this case again the quark-initiated jets are simpler to evaluate 
than the gluon-initiated ones. 

Our present target is to evaluate the NLO contributions to the production of a 
forward jet this time coupled to a bound state of two reggeized gluons, or hard 
pomeron, which lives in a color singlet representation and does not have associated 
minijet radiation but a rapidity gap instead. The calculation is also divided into quark and gluon initiated jets and has numerous applications in the field of diffractive jet production, the most famous one being Mueller-Tang jet production \cite{tang}. In this observable two jets are emitted in the forward direction of each hadron and a large rapidity gap sits in between them. 

Technically,  this calculation is rather involved and in this letter we present the final results to be used for phenomenology. In separate publications \cite{toappear} we will show the 
detailed evaluation of both the quark and gluon initiated components. We first make 
a nutshell presentation of the high energy effective action indicating the relevant 
effective Feynman rules which are needed in our calculation. Then we introduce 
the notation to understand the relevant variables present in the description of the 
jet vertex, to finally describe our results and discuss directions for future  
theoretical and phenomenological studies.

\section{Basics of the high energy effective action}
We are interested in dijet production in hadron-hadron collisions at very high energies, $p (p_A)+p (p_B) \to J_1(p_{J,1})+J_2(p_{J,2})+{\rm gap}$, where there exists a large region in rapidity $\Delta y_{\rm gap}$ in between the tagged jets without hadronic activity. In perturbative QCD we understand this rapidity gap as generated by a color singlet exchange in the $t$-channel which takes the form of a hard or BFKL pomeron whose interactions with external particles is well described by Lipatov's high energy effective action~\cite{effective}. To the usual 
QCD action we add an extra piece which accounts for the interaction of reggeized $t$-channel gluons (reggeons) and normal quarks and gluons: $S_{\rm eff}=S_{\rm QCD}+S_{\rm ind}$. The extra ``induced" piece reads
\begin{equation}
S_{\rm ind}=\int {\rm d}^4 x\,{\rm Tr}[(W_-[v(x)]-A_-(x))\partial_\perp^2 A_+(x)+\{+\leftrightarrow -\}],
\end{equation}
where $v_\mu=-iT^av_\mu^a(x)$ is the gluon field, and $A_\pm(x)=-iT^aA^a_\pm(x)$ is the reggeon field, introduced as a new degree of freedom, which mediates any interaction between clusters of emitted particles well  separated in rapidity. 

\begin{figure}[htb]
    \label{fig:subfigures}
   \centering
   \parbox{.7cm}{\includegraphics[height = 1.8cm]{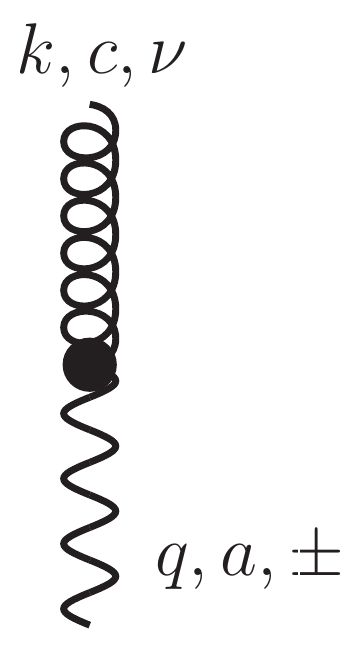}} $=  \displaystyle 
   \begin{array}[h]{ll}
    \\  \\ - i{\bm q}^2 \delta^{a c} (n^\pm)^\nu,  \\ \\  \qquad   k^\pm = 0.
   \end{array}  $ 
 \parbox{1.2cm}{ \includegraphics[height = 1.8cm]{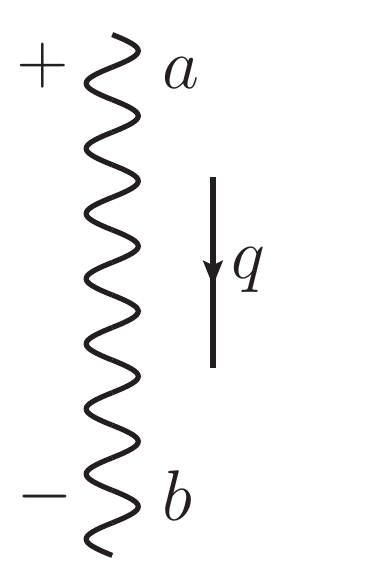}}  $=  \displaystyle    \begin{array}[h]{ll}
    \delta^{ab} \frac{ i/2}{{\bm q}^2} \end{array}$ \\
 \parbox{1.7cm}{\includegraphics[height = 1.8cm]{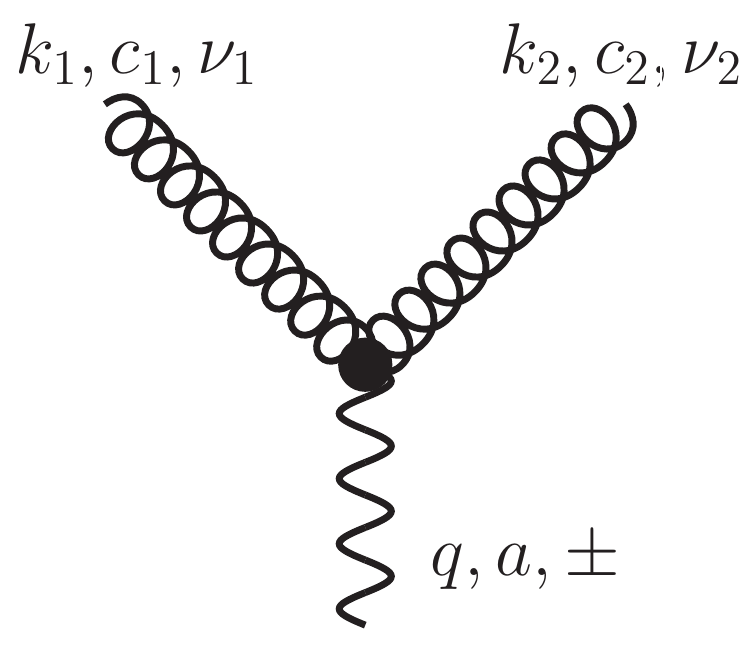}} $ \displaystyle  =  \begin{array}[h]{ll}  \\ \\ g f^{c_1 c_2 a} \frac{{\bm q}^2}{k_1^\pm}   (n^\pm)^{\nu_1} (n^\pm)^{\nu_2},  \quad  k_1^\pm  + k_2^\pm  = 0
 \end{array}$\\
  \parbox{2.8cm}{\includegraphics[height = 1.8cm]{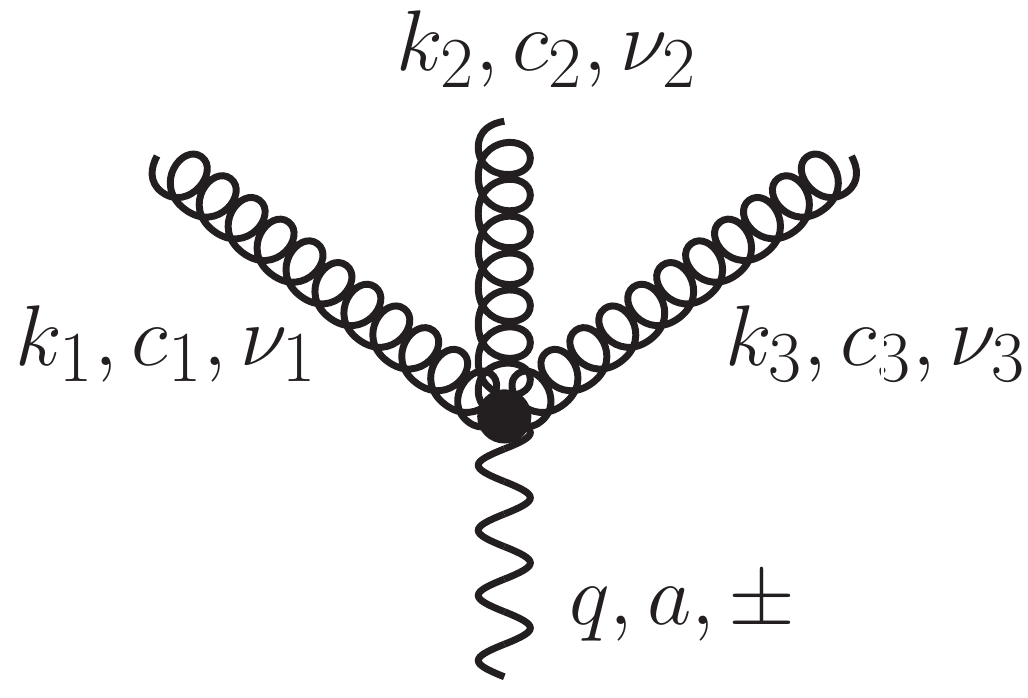}} $ \displaystyle  =  \begin{array}[h]{ll}  \\ \\ ig^2\bm{q}^2 \left(\frac{f^{c_3c_2c}f^{c_1ca}}{k_3^\pm k_1^\pm}+\frac{f^{c_3c_1c}f^{c_2ca}}{k_3^\pm k_2^\pm}\right)   (n^\pm)^{\nu_1} (n^\pm)^{\nu_2}(n^\pm)^{\nu_3},  \\ \\ \quad  k_1^\pm  + k_2^\pm + k_3^\pm = 0
 \end{array}$
 \caption{\small Feynman rules for the lowest-order effective vertices of the effective action \cite{antonov}. Wavy lines denote reggeized fields and curly lines gluons. Pole prescriptions for the light-cone denominators are discussed in \cite{pole}.}
\label{fig:3}
\end{figure}

Locally in rapidity, within each cluster, the reggeon-gluon interactions are mediated by the Wilson line couplings 
\begin{eqnarray}
W_\pm [v(x)]=-\frac{1}{g}\partial_\pm {\cal P}\exp \left\{-\frac{g}{2}\int_{-\infty}^{x^\mp} \rmd z^\pm v_\pm (z)\right\}.
\end{eqnarray}
The reggeon field satisfies the kinematic constraint $\partial_\pm A_\mp (x)$ $=0$ which is present in the  Feynman rules of Fig. \ref{fig:3}. We use the Sudakov decomposition $k=k^+ n^- / 2 + k^- n^+ / 2 +\bm{k}$ where 
$n^\pm=2p_{A,B}/\sqrt{s}$ and the squared center-of-mass energy is $s=2p_A\cdot p_B$. 
For the hadrons we write $p_A = p_A^+ n^- / 2$ and $p_B = p_B^-  n^+ / 2$, while
 for the jets
$p_{J,i} = \sqrt{{\bm k}_{J,i}^2}  \left(  e^{y_{J,i}} n^- /2 +   e^{-y_{J,i}} n^+ /2 \right) + {\bm k}_{J,i},\,i=1,2$. $\bm{k}_{J,i}$ and $y_{J,i}$ are the transverse momenta and rapidity of the jets.
To understand the general idea of our calculation let us point out that at LO parton level we 
consider the high energy limit of the process  $i(p_a)+j(p_b)\to k(p_1)+l(p_2)$, with  $i,j,k, l = q, \bar{q}, g$ and  
a color singlet exchange in the $t$-channel. This requires the exchange of at least two reggeized gluons. The corresponding partonic LO cross-section 
is
\begin{equation}\label{kreutz}
\frac{ {\rm d}\hat{\sigma}_{ij}}{ {\rm d}^{2}\bm{k}}=\int {\rm d}^{2}\bm{l}_1 
{\rm d}^{2}\bm{l}_2
\frac{h_{i,{\rm a}}^{(0)}}{\pi \bm{l}_1^2(\bm{k}-\bm{l}_1)^2}
\frac{h_{j,{\rm b}}^{(0)}}{\pi \bm{l}_2^2(\bm{k}-\bm{l}_2)^2},
\end{equation}
where $h^{(0)}_i$, $i = g, q, \bar{q}$ is the parton - two reggeized gluon vertex at LO and $\bm{k}$ the 
momentum transfer. Resumming $\Delta y_{\rm
  gap} \sim \ln{s/s_0}$ terms we have
\begin{align}\label{resum}
\frac{ \rmd\hat{\sigma}_{ij}}{\rmd^2 {\bm k}} &= \int\frac{{\rm d}^2\bm{l}_1 {\rm d}^2\bm{l}_1'}{\pi} 
\frac{{\rm d}^2\bm{l}_2 {\rm d}^2\bm{l}_2'}{\pi}
h_{i,{\rm a}}^{(0)}
h_{j,{\rm b}}^{(0)}
 G \left(\bm{l}_1,\bm{l}_1',\bm{k},\frac{s}{s_0}\right) 
G 	\left(\bm{l}_2,\bm{l}_2',\bm{k},\frac{s}{s_0} \right),
\end{align}
where $G$ is the non-forward BFKL Green function \cite{nonforward}.
In the following we determine the NLO corrections to the parton-2
reggeized gluon couplings. They include the one-loop virtual
corrections to the tree-level amplitude, already computed in
\cite{fadin}. The inelastic processes to be included in our
calculation require the additional emission of a gluon (gluon and
(anti-)quark initated case) and splitting of the gluon into into
quark-antiquark pair (gluon initiated case) in the forward region of
the initial state partons.  We discuss them in some detail in the
following.

\section{Real NLO Corrections to Impact Factors}
\begin{figure}[t]
  \centering
\includegraphics[width = 3cm]{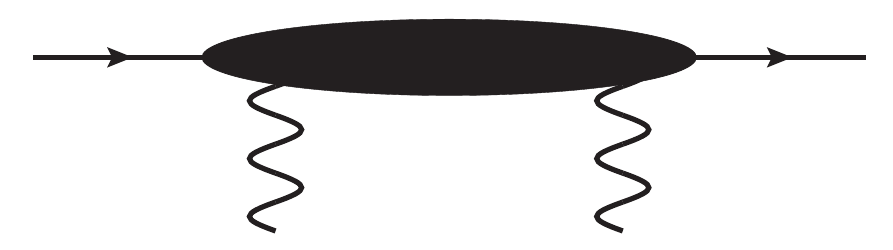} =  
\includegraphics[width = 3cm]{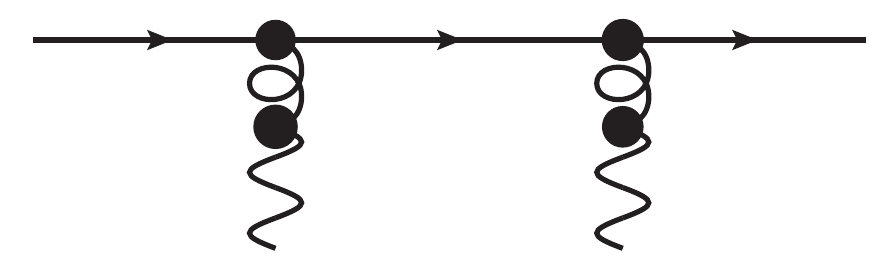} + \includegraphics[width = 3cm]{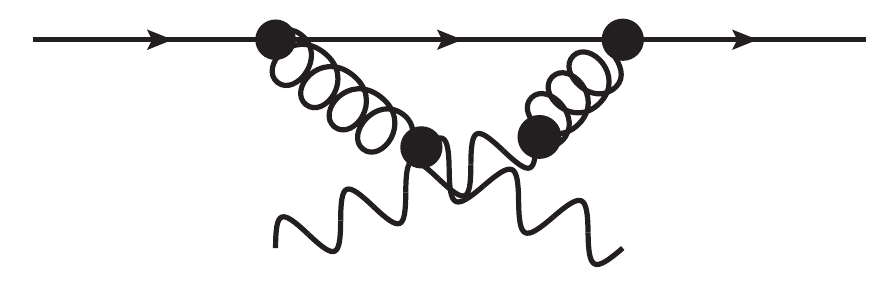}
  \caption{\it Diagrams for the LO impact factor $h^{(0)}$ for quark-initiated jets. With the definitions \eqref{definitions}, the leading order impact factors read $h_q^{(0)}=C_f^2h^{(0)}$ and $h_g^{(0)}=C_a^2(1+\epsilon)h^{(0)}$.}
  \label{fig:schematic}
\end{figure}
\begin{figure}[t]
\centering
\includegraphics[scale=1.2]{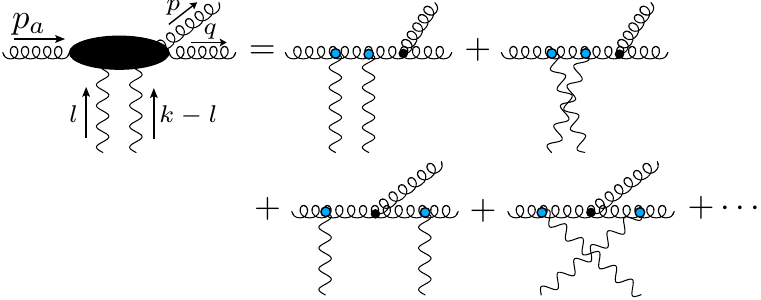}
\caption{\it Effective vertex in the quasielastic corrections to gluon-initiated jets with $gg$ final state. 
}
\label{sample}
\end{figure}

In our framework the typical diagrams to be evaluated are shown in Figs.~\ref{fig:schematic} and \ref{sample}. We 
work in Sudakov variables for the momenta in Fig.~\ref{sample}: $p_a=p_a^+ n^- /2$, $k=k^- n^+ / 2 +\bm{k}$, $l=l^-  n^+ / 2 +\bm{l}$, $p=(1-z)p_a^+ n^- / 2 + \bm{p}^2 n^+ / (2 (1-z)p_a^+) +\bm{p}$, 
$q=zp_a^+ n^- / 2 + \bm{q}^2  n^+ / (2 z p_a^+) +\bm{q}$. In the integration over the reggeon loop momentum 
we incorporate its longitudinal component into the definition of the impact factors. With the notation
\begin{equation}
i\{\phi_{qqg},\,\phi_{ggg},\,\phi_{gq\bar{q}}\}=\int\frac{{\rm d}l^-}{8\pi}\{i{\cal M}^{cde}_{q2r^*\to qg},\,i{\cal M}^{abcde}_{g2r^*\to gg},\,i{\cal M}^{ade}_{g2r^*\to q\bar{q}}\}P^{de},
\end{equation}
where $P^{de}=\delta^{de}/\sqrt{N_c^2-1}$ is the projector onto color singlet for the two-reggeon state, we can express the real contribution to the NLO correction to the impact factor, $h^{(1)}$, appearing in \eqref{resum} as
\begin{equation}
h^{(1)}_{r,\,X,\,{\rm a}} \rmd\Gamma^{(2)} =  \frac{2^\epsilon}{(4\pi)^{4+3\epsilon}(p_a^+)^2} 
\frac{\overline{|\phi_{X,\,{\rm a}}|^2}}{z(1-z)} \rmd\Gamma^{(2)},\quad _{X=\{qqg,\,ggg,\,qq\bar{q}\}}.
\end{equation} 
with $\rmd \Gamma^{(2)} =  \rmd z \rmd^{2+2\epsilon}\bm{q} / \pi^{1+\epsilon}$.  Associated to each forward jet, there exists a diffractive system in the forward region of the corresponding hadron, with momentum transfer $t = -{\bm k}^2$ and an invariant mass $M_X$. The upper limit on the latter at  parton level stems from the 
constraint  $\hat{M}_X^2 = (p_a + k)^2 < \hat{M}_{X, \rm max}^2 = x {M}_{X, \rm max}^2 + (1-x) t $ 
where $M_X < M_{X, \rm max}$ is limited by experiment. We therefore find
\begin{equation}
\begin{aligned}
& h^{(1)}_{r,\,X} \rmd \Gamma^{(2)} =\frac{h^{(0)} (1 + \epsilon)}{\mu^{2\epsilon}\Gamma(1-\epsilon)}\frac{\alpha_{s,\epsilon}}{2\pi}\,\Theta\left(\hat{M}_{X,\,{\rm max}}^2-\frac{\bm{\Delta}}{z(1-z)}\right)
\\&
\left\{ \tfrac{1}{1 + \epsilon}P_{gq}(z,\epsilon),\tfrac{1}{2!}P_{gg}(z, \epsilon), P_{qg}(z,\epsilon)\right\}
J_{X}(\bm{q},\bm{k},\bm{l}_1,\bm{l}_2,z) \rmd \Gamma^{(2)},
\end{aligned}
\end{equation}
 where $P_{gq}(z,\epsilon)=C_f\frac{1+(1-z)^2+\epsilon z^2}{z},\,P_{gg}(z, \epsilon)=2C_a\frac{(1-z(1-z))^2}{z(1-z)}$ and $P_{qg}(z,\epsilon)=\frac{1}{2}\left(1-\frac{2z(1-z)}{1+\epsilon}\right)$ are the real part of the Altarelli-Parisi splitting functions in $d=4+2\epsilon$ dimensions and we defined
\begin{align}\label{definitions}
&\alpha_{s,\epsilon}=\frac{g^2\mu^{2\epsilon}\Gamma(1-\epsilon)}{(4\pi)^{1+\epsilon}},\qquad h^{(0)}=\frac{\alpha_{s,\epsilon}^2 2^\epsilon}{\mu^{4\epsilon}\Gamma^2(1-\epsilon)(N_c^2-1)},\notag\\&\bm{\Delta}=\bm{q}-z\bm{k},\quad\bm{\Sigma}_i=\bm{q}-\bm{l}_i,\quad\bm{\Upsilon}_i=\bm{q}-\bm{k}+\bm{l}_i,\quad i=1,2,\notag\\&
J_{X}(\bm{q},\bm{k},\bm{l}_1,\bm{l}_2,z)=\Bigg[\{C_f,C_a,C_a\}\frac{\bm{\Delta}}{\bm{\Delta}^2}-\{C_f,C_a,C_f\}\frac{\bm{q}}{\bm{q}^2}
\\
& -   \{C_a,C_a,C_f\}\frac{\bm{p}}{\bm{p}^2}  -   
  \frac{1}{2}\{C_a,C_a,-\frac{1}{C_a}\}  \left(\frac{\bm{\Sigma}_1}{\bm{\Sigma}_1^2}
+
\frac{\bm{\Upsilon}_1}{\bm{\Upsilon}_1^2}\right)\Bigg]\cdot \Bigg[\{1\leftrightarrow 2\}\Bigg].\notag
\end{align}
\section{The NLO Mueller-Tang Jet Vertex}
In order to define an infrared and collinear safe NLO  cross section, we need to convolute the partonic cross 
section with a jet function $S_J$:
\begin{equation}
\frac{\rmd\hat{\sigma}_J}{\rmd J_1\rmd J_2\rmd^2\bm{k}}=\rmd\hat{\sigma}\otimes S_{J_1}S_{J_2},\quad \rmd J_i=\rmd^{2+2\epsilon}\bm{k}_{J_i}\rmd y_{J_i},\,i=1,2.
\end{equation}
Infrared finiteness imposes general constraints on the jet function  \cite{seymour}. For two final state partons, the  jet function $S_J^{(3)}(\bm{p},\bm{q},zx,x)$ must be $ \{ \bm{q}, z \} \leftrightarrow \{\bm{p}, 1-z\}$ symmetric, and must reduce to the one final state parton distribution  $S_{J}^{(2)}(\bm{p},x)=x\,\delta\left(x-\frac{|\bm{k}_J|e^{y_J}}{\sqrt{s}}\right)\delta^{2+2\epsilon}(\bm{p}-\bm{k}_J)$ in the soft and collinear limits. In particular
\begin{equation}\label{propjet}
S^{(3)}_J(\bm{p},\bm{q},zx,x)\stackrel{\bm{p}\to 0}{\longrightarrow}S^{(2)}_J(\bm{k},zx);\quad S^{(3)}_J(\bm{p},\bm{q},zx,x)\stackrel{\tfrac{\bm{q}}{z}\to\tfrac{\bm{p}}{1-z}}{\longrightarrow}S_J^{(2)}(\bm{k},x).
\end{equation}
Completing our result with the virtual  corrections calculated in \cite{fadin}, taking into account  UV renormalization of the QCD Lagrangian, and absorbing initial state collinear emissions into a redefinition of parton distribution functions, we obtain,  within the collinear factorization framework, the result
\begin{equation}
\begin{aligned}
\hspace{-0.25cm}\frac{\rmd\sigma_{J,H_1H_2}}{\rmd J_1\rmd J_2\rmd^2\bm{k}}&=\frac{1}{\pi^2}\int\rmd\bm{l}_1\rmd\bm{l}_1'\rmd\bm{l}_2\rmd\bm{l}_2'\frac{\rmd V(\bm{l}_1,\bm{l}_2,\bm{k},\bm{p}_{J,1},y_1,s_0)}{\rmd J_1}
\\\times
&
 G\left(\bm{l}_1,\bm{l}_1',\bm{k}, \frac{\hat{s}}{s_0}\right)
G\left(\bm{l}_2,\bm{l}_2',\bm{k},\frac{\hat{s}}{s_0} \right)\frac{\rmd V(\bm{l}_1',\bm{l}_2',\bm{k},\bm{p}_{J,2},y_2,s_0)}{\rmd J_2},
\end{aligned}
\end{equation}
where $\hat{s}=x_1x_2s$, $x_0= -t / (M_{x, \rm max}^2 - t)$ and
\begin{align}\label{jetv}
&\frac{\rmd V}{\rmd J}=\sum_{j=\{q_k,\bar{q}_k,g\}}^{k=1,\cdots,n_f}\int_{x_0}^1\rmd x\,f_{j/H}(x,\mu_F^2)
\left(\tfrac{\rmd \hat{V}^{(0)}_j}{\rmd J}+\tfrac{\rmd \hat{V}^{(1)}_j}{\rmd J}\right),
\, \, 
\frac{\rmd \hat{V}_j^{(0)}}{\rmd J} = \frac{\alpha_s^2 C_j^2 }{N_c^2 -1}S_J^{(2)}({\bm k}, x),
\notag\\
&
\frac{\rmd \hat{V}_j^{(1)}}{\rmd J}=
\int d\Gamma^{(2)} \left(\frac{\rmd \hat{V}_{j,\,v}^{(1)}}{\rmd J}+\frac{\rmd \hat{V}_{j,\,r}^{(1)}}{\rmd J}+\frac{\rmd \hat{V}_{j,\,{\rm UV\,ct.}}^{(1)}}{\rmd J}
+\frac{\rmd \hat{V}_{j,\,{\rm col.\,ct.}}^{(1)}}{\rmd J} \right),\notag\\
&\frac{\rmd\hat{V}^{(1)}_{r,\,\{q_k/\bar{q}_k,g\}}}{\rmd J}=\left\{h^{(1)}_{r,\,qqg}, h^{(1)}_{r,\,q\bar{q}g}+h^{(1)}_{r,\,ggg}\right\}S^{(3)}_J(\bm{p},\bm{q},zx,x),\\&\tfrac{\rmd \hat{V}_{\{q_k/\bar{q}_k,g\},\,{\rm UV\,ct.}}^{(1)}}{\rmd J}=\{h_q^{(0)},h_g^{(0)}\}\frac{\alpha_{s,\epsilon}}{2\pi}\frac{\beta_0}{\epsilon}S^{(2)}_J(\bm{k},x),~~\tfrac{\rmd\hat{V}_{\{g,q/\bar{q}\}}^{(0)}}{\rmd J}=h^{(0)}_{\{g,q\}}S_J^{(2)}(\bm{k},x),\notag\\&\tfrac{\rmd \hat{V}_{j,\,{\rm col.\,ct.}}^{(1)}}{\rmd J}=-\frac{\alpha_{s,\epsilon}}{2\pi}
\left(\tfrac{1}{\epsilon}+\ln\tfrac{\mu_F^2}{\mu^2}\right)
\int_0^1\rmd z\,S_J^{(2)}(\bm{k},zx)\sum_{i=\{q_\ell,\bar{q}_\ell,g\}}^{\ell=1,\cdots,n_f}h_i^{(0)}P_{ij}^{(0)}(z),\notag
\end{align}
with $\beta_0=\frac{11}{3}C_a-\frac{2}{3}n_f$, $P_{ij}^{(0)}(z)$ the
LO DGLAP splitting functions and $C_{q,\bar{q}} = C_f, C_g = C_a$. The result for $\frac{\rmd
  \hat{V}^{(1)}_{j, v}}{\rmd J}$ can be extracted from \cite{fadin}.
Note that, when expanded to NLO, our result is independent of the scale $s_0$.

To write a physical representation of this vertex in dimension four we introduce a phase slicing parameter, 
$\lambda^2\ll\bm{k}^2$, to regularize the singular regions in phase space. Using  the limits in  
Eq.~\eqref{propjet} we can rewrite $\rmd V_{q,g} / \rmd J$ in terms of $\lambda$~\cite{toappear} and, 
introducing the notations ($i=1,2$)
{\small\begin{align}
\label{pepe}
     P_{0}(z) =C_a&\big[\tfrac{2(1-z)}{z}+z(1-z)\big],\quad P_{1}(z)=C_a\big[\tfrac{2z}{[1-z]_+}+z(1-z)\big]  ,\notag\\
P_{qq}^{(0)}(z)  = C_f& \left(\frac{1+z^2}{1-z} \right)_+, \quad P_{qg}^{(0)(z)}= \frac{z^2 + (1-z)^2}{2} \; , \notag \\
 P^{(0)}_{gq}(z)= C_f& \frac{1 + (1-z)^2}{z}, ~~ P_{gg}^{(0)}(z) = P_0(z) + P_1(z) + \frac{\beta_0}{2}\delta(1-z) \; ,
\notag \\
    \alpha_s= \alpha_s (\mu^2) & , \qquad \phi_i=\arccos\tfrac{ \bm{l}_i \cdot ( \bm{k} - {\bm l}_i)}{|\bm{l}_i||\bm{k}-\bm{l}_i|}, \notag\\
    J_{1} ({\bm q}, {\bm k}, {\bm l}_i, z) & = \frac{1}{4} \bigg[ 2
    \frac{{\bm k}^2}{{\bm p}^2} \bigg(\frac{(1-z)^2}{{\bm \Delta}^2} -
    \frac{1}{{\bm q}^2} \bigg) - \frac{1}{{\bm \Sigma}_i^2} \bigg(
    \frac{({\bm l}_i - z {\bm k})^2}{{\bm \Delta}^2} - \frac{{\bm
        l}_i^2}{{\bm q}^2}
    \bigg)\notag \\
    & \qquad - \frac{1}{{\bm \Upsilon}_i^2} \bigg( \frac{({\bm l}_i -
      (1-z) {\bm k})^2}{{\bm \Delta}^2} - \frac{({\bm l}_i - {\bm
        k})^2}{{\bm q}^2} \bigg)
    \bigg],~i=1,2;  \notag\\
    J_{2} ({\bm q}, {\bm k}, {\bm l}_1, {\bm l}_2) &= \frac{1}{4}
    \bigg[ \frac{{\bm l}_1^2}{ {\bm p}^2 {\bm \Upsilon}^2_1} + \frac{(
      {\bm k} - {\bm l}_1)^2}{ {\bm p}^2 {\bm \Sigma}^2_1} +
    \frac{{\bm l}_2^2}{ {\bm p}^2 {\bm \Upsilon}^2_2} + \frac{( {\bm
        k} - {\bm l}_2)^2}{ {\bm p}^2 {\bm \Sigma}^2_2}
    \notag\\
    & \hspace{-1cm}- \frac{1}{2} \bigg( \frac{({\bm l}_1 - {\bm
        l}_2)^2}{{\bm \Sigma}_1^2 {\bm \Sigma}_2^2} + \frac{({\bm k} -
      {\bm l}_1 - {\bm l}_2)^2}{ {\bm \Upsilon}_1^2 {\bm \Sigma}_2^2 }
    + \frac{({\bm k} - {\bm l}_1 - {\bm l}_2)^2}{ {\bm \Sigma}_1^2
      {\bm \Upsilon}_2^2 } + \frac{({\bm l}_1 - {\bm l}_2)^2}{{\bm
        \Upsilon}_1^2 {\bm \Upsilon}_2^2} \bigg)
    \bigg],
\end{align}}
we present our expression for those jets with a quark as the initial state, {\it i.e.}
{\small\begin{align}
 &   \frac{\rmd\hat{V}^{(1)}_q(x, {\bm k}, {\bm l}_1, {\bm l}_2; x_J, {\bm k}_J; M_{X,\text{max}}, s_0)}{\rmd J}  =   v^{(0)}\frac{\alpha_s}{2 \pi}  \big(Q_1 + Q_2 + Q_3 \big)
\notag \\
& Q_1 =   S_J^{(2)}({\bm k}, x) C_f^2 { \Bigg[-\frac{\beta_0}{4}
\bigg\{\left[\ln\left(\frac{\bm{l}_1^2}{\mu^2}\right)+\ln\left(\frac{(\bm{l}_1-\bm{k})^2}{\mu^2}\right)+\{1\leftrightarrow 2\}\right]} 
\notag \\ &  -\frac{20}{3}\bigg\} -4C_f + \frac{C_a}{2}\bigg(\bigg\{
\frac{3}{2\bm{k}^2}
\bigg[
\bm{l}_1^2\ln\left(\frac{(\bm{l}_1-\bm{k})^2}{\bm{l}_1^2}\right)+(\bm{l}_1-\bm{k})^2 \; \cdot
\notag \\
& \ln\left(\frac{\bm{l}_1^2}{(\bm{l}_1-\bm{k})^2}\right)  -4|\bm{l}_1||\bm{l}_1-\bm{k}|\phi_1\sin\phi_1\bigg]
  -\frac{3}{2}\bigg[\ln\left(\frac{\bm{l}_1^2}{\bm{k}^2}\right)
 \notag \\
&  +\ln\left(\frac{(\bm{l}_1-\bm{k})^2}{\bm{k}^2}\right)\bigg]
 -\ln\left(\frac{\bm{l}_1^2}{\bm{k}^2}\right)\ln\left(\frac{(\bm{l}_1-\bm{k})^2}{s_0}\right)
-\ln\left(\frac{(\bm{l}_1-\bm{k})^2}{\bm{k}^2}\right) \; \cdot
\notag \\
&
\ln\left(\frac{\bm{l}_1^2}{s_0}\right) -2\phi_1^2+\{1\leftrightarrow 2\}\bigg\} +2\pi^2+\frac{14}{3}\bigg)\Bigg]\;,
\notag \\
& Q_2 =
\int_{z_0}^1 \rmd z  \,\, 
 S_J^{(2)}({\bm k}, z x)%
\bigg[ \ln \frac{\lambda^2}{\mu^2_F}   \left(  C_f^2
P^{(0)}_{qq}(z)+ {C_a^2}P^{(0)}_{gq}(z) \right) 
 \notag \\
&
+ C_f (1-z)\left(C_f^2 -\frac{2}{z} {C_a^2}\right)
+2 C_f (1+z^2)\left(\frac{\ln(1-z)}{1-z}\right)_+\bigg] \; ,\notag
\\ 
& Q_3 = \int_0^1 \rmd z\int\frac{\rmd^2\bm{q}}{\pi} \bigg[
 \Theta\left(\hat{ M}^2_{X,{\rm max}}-\frac{({ \bm p } - z {\bm k})^2}{z(1-z)}\right) S_J^{(3)}(\bm{p},\bm{q},(1-z)x,x)  C_f^2  
\notag \\
&
  P^{(0)}_{qq}(z)\Theta\left(\frac{|\bm{q}|}{1-z}-\lambda^2\right)  
{ \frac{  \bm{k}^2}{\bm{q}^2(\bm{p}-z\bm{k})^2}}
 + \Theta\left(\hat{M}_{X, \rm max}^2 - \frac{{\bm \Delta}^2}{z(1-z)} \right)
 \notag \\
&  S_J^{(3)}(\bm{p},\bm{q},zx,x)  
   P_{gq}^{(0)}(z) 
 \big\{C_f C_a [J_1(\bm{q},\bm{k},\bm{l}_1)+J_1(\bm{q},\bm{k},\bm{l}_2)]
\notag \\ &
\hspace{6cm}
+ {C_a^2} J_2(\bm{q},\bm{k},\bm{l}_1,\bm{l}_2) \Theta(\bm{p}^2-\lambda^2)
    \big\}\bigg] \; .
\end{align}}
\newpage
In a similar way, the equivalent gluon-generated forward jet vertex reads
{\small\begin{align}\label{apocalipsis}
&\frac{\rmd\hat{V}^{(1)}(x,\bm{k},\bm{l}_1,\bm{l}_2;x_J,\bm{k}_J;M_{X,\,{\rm max}},s_0)}{\rmd J}  =v^{(0)}\frac{\alpha_s}{2\pi}\big( G_1 + G_2 + G_3 \Big)
\notag \\
& G_1  =
C_a^2\,S_J^{(2)}(\bm{k},x)
\Bigg[C_a \left( \pi^2- \frac{5}{6} \right)- \beta_0 \left( \ln\frac{\lambda^2}{\mu^2} -  \frac{4}{3} \right) 
\notag \\
&
+\left(\frac{\beta_0}{4} + \frac{11 C_a}{12}+\frac{n_f}{6 C_a^2}\right)
\left( \ln\frac{\bm{k}^4}{ {\bm l}_1^2  ({\bm k} - {\bm l}_1^2)}
+
 \ln\frac{\bm{k}^4}{ {\bm l}_2^2  ({\bm k} - {\bm l}_2)^2 } 
\right)
\notag \\
&
+\frac{1}{2}\bigg\{C_a\bigg(
\ln^2\frac{\bm{l}_1^2}{(\bm{k}-\bm{l}_1)^2}
+
\ln\frac{\bm{k}^2}{\bm{l}_1^2}\ln\frac{\bm{l}_1^2}{s_0}
+
\ln\frac{\bm{k}^2}{(\bm{k}-\bm{l}_1)^2}\ln\frac{(\bm{k}-\bm{l}_1)^2}{s_0}
\bigg)
\notag\\
&-\bigg(\frac{n_f}{3 C_a^2} + \frac{11 C_a}{6}\bigg)
\frac{\bm{l}_1^2-(\bm{k}-\bm{l}_1)^2}{\bm{k}^2}
\ln\frac{\bm{l}_1^2}{(\bm{k}-\bm{l}_1)^2}
-2\bigg(\frac{n_f}{C_a^2}+4C_a\bigg)
\notag\\
&  \frac{(\bm{l}_1^2(\bm{k}-\bm{l}_1)^2)^{\frac{1}{2}}}{\bm{k}^2} \phi_1\sin\phi_1 
+\frac{1}{3}\bigg(
C_a+\frac{n_f}{C_a^2}
\bigg)\bigg[
16\frac{(\bm{l}_1^2(\bm{k}-\bm{l}_1)^2)^{\frac{3}{2}}}{(\bm{k}^2)^3}\phi_1\sin^3\phi_1
\notag\\
& -4\frac{\bm{l}_1^2(\bm{k}-\bm{l}_1)^2}{(\bm{k}^2)^2}  \bigg(
2-\frac{\bm{l}_1^2-(\bm{k}-\bm{l}_1)^2}{\bm{k}^2}\ln\frac{\bm{l}_1^2}{(\bm{k}-\bm{l}_1)^2}
\bigg)
\sin^2\phi_1+\frac{(\bm{l}_1^2(\bm{k}-\bm{l}_1)^2)^{\frac{1}{2}}}{(\bm{k}^2)^2}
\notag\\ & \cos\phi_1
\bigg(4\bm{k}^2 -12(\bm{l}_1^2(\bm{k}-\bm{l}_1)^2)^{\frac{1}{2}}\phi_1\sin\phi_1-(\bm{l}_1^2-(\bm{k}-\bm{l}_1)^2)\ln\frac{\bm{l}_1^2}{(\bm{k}-\bm{l}_1)^2}\bigg)
\bigg]
\notag\\&-2C_a\phi_1^2+\{\bm{l}_1\leftrightarrow\bm{l}_2,\phi_1\leftrightarrow\phi_2\}\bigg\}\Bigg]
\notag \\
 & G_2 =   \int_{z_0}^1\rmd z\,S_J^{(2)}(\bm{k},zx) \bigg\{  2n_f P_{qg}^{(0)}(z)
\left(C_f^2 \ln \frac{\lambda^2}{\mu_F^2}  +  C_a^2 \ln(1-z) \right) 
\notag  
 \\ &
 + C_a^2 P_{gg}^{(0)}(z) \ln\frac{\lambda^2}{\mu_F^2}  + C_f^2 n_f + 2 C_a^3 z \bigg(  (1-z)\ln(1-z)   + 2 \left[\frac{\ln (1-z)}{1-z} \right]_+ \bigg)
\notag \\
&G_3 = 
\int_0^1 \rmd z\int\frac{\rmd^2\bm{q}}{\pi} \bigg\{
n_f P^{(0)}_{qg}(z) 
\bigg[
C_a^2 \Theta\left(\hat{M}_{X,{\rm max}}^2-\frac{z {\bm p}^2}{(1-z)}\right)
\notag \\
&
    S^{(3)}_J(\bm{k}-z\bm{q},z\bm{q},zx,x) \bigg[ \frac{\Theta({\bm p^2 - \lambda^2}) {\bm k}^2}{({\bm p}^2 + {\bm q}^2) {\bm p}^2} +
\frac{ {\bm k}^2}{({\bm p}^2 + {\bm q}^2) {\bm q}^2} \bigg]
\notag \\
&
- \Theta\left(\hat{M}_{X,{\rm max}}^2-\frac{{\bm \Delta}^2}{z(1-z)}\right) 
  S^{(3)}_J(\bm{p},\bm{q}, zx, x)
\bigg(C_a^2  \frac{ {\bm k}^2}{({\bm p}^2 + {\bm q}^2) {\bm q}^2} 
\notag \\
&
-   
 2 C_f^2 
 \frac{{\bm k}^2 \Theta({\bm q}^2 - \lambda^2)}{({\bm p}^2 + {\bm q}^2) {\bm q}^2  } \bigg)\bigg]
+ P_1(z)  \Theta\left(\hat{ M}^2_{X,{\rm max}}-\frac{({ \bm p } - z {\bm k})^2}{z(1-z)}\right) 
\notag \\ & 
 S_J^{(3)}({\bm p}, {\bm q}, (1-z)x, x) \frac{(1-z)^2 {\bm k}^2}{(1-z)^2 ({\bm p} - z {\bm k})^2 + {\bm q}^2}
\bigg[  \Theta\left(\frac{|\bm{q}|}{1-z}-\lambda\right)  \frac{1}{{\bm q}^2} 
\notag \\
 &+ 
\Theta\left(\frac{|\bm{p} - z {\bm k}|}{1-z}-\lambda\right)  \frac{1}{({\bm p} - z {\bm k})^2}
 + 
\Theta\left(\hat{M}_{X, \rm max}^2 - \frac{{\bm \Delta}^2}{z(1-z)} \right)
 S_J^{(3)}(\bm{p},\bm{q},zx,x) 
& \notag \\
 &
\bigg[
\frac{n_f }{ C_a^2}  P_{qg}^{(0)} \bigg(J_2(\bm{q},\bm{k},\bm{l}_1,\bm{l}_2)   - \frac{{\bm k}^2}{{\bm p}^2( {\bm q}^2 + {\bm p}^2)}\bigg)
- n_f  P_{qg}^{(0)} \bigg(
J_{1} ({\bm q}, {\bm k}, {\bm l}_1, z)  \notag \\
&  + J_{1} ({\bm q}, {\bm k}, {\bm l}_2, z)  
 \bigg) 
+   P_0(z) 
 \bigg( J_1(\bm{q},\bm{k},\bm{l}_1)+J_1(\bm{q},\bm{k},\bm{l}_2) \notag \\
& \hspace{7cm} + J_2(\bm{q},\bm{k},\bm{l}_1,\bm{l}_2) \Theta(\bm{p}^2-\lambda^2)
    \bigg)\bigg] 
\bigg\}.
\end{align}}
These expressions, although lengthy, are suited to perform phenomenological studies. It is important to note that its 
convolution with the nonforward BFKL Green function with exact treatment of the running of the coupling 
is complicated and Monte Carlo integration techniques~\cite{Chachamis:2011nz} are required in order to generate exclusive distributions needed to 
describe different diffractive data in hadronic collisions, in particular those already recorded at the LHC.

\section{Outlook}
In this brief letter we have presented our final results for the jet vertex describing the 
coupling of a hard pomeron to a forward jet with a next-to-leading order accuracy. This 
result is a necessary step towards the phenomenological study of diffractive jet production 
at high energies. Together with the nonforward gluon Green function it allows for the study 
of different diffractive cross-sections with great detail, in particular when the latter is implemented in a Monte Carlo event generator~\cite{Chachamis:2011nz}. Running coupling, energy scale choice 
and renormalization scheme issues can be addressed fully at NLO, greatly increasing the precision of our predictions. 

Our procedure to use the high energy effective action proposed by Lipatov can be extended to other observables, in particular those where non-linear effects might be 
important. These non-linearities can be relevant already at the Large Hadron Collider and 
would definitely play an important role at possible future experiments such as the 
Large Hadron electron Collider~\cite{AbelleiraFernandez:2012cc}. 
\\
\\
\\
{\small 
{\bf  Acknowledgements:} We thank the participants of the $2^{\rm nd}$ {\em Informal Meeting on Scattering Amplitudes \& the Multi-Regge Limit} (Madrid, February 2014), for stimulating discussions.  MH acknowledges support from U.S. Department of Energy (DE-AC02-98CH10886) and ``BNL Laboratory Directed Research \& 
Development'' grant (LDRD 12-034).  JDM is supported by European Research Council under Advanced Investigator Grant ERC-AD-267258. ASV acknowledges support from European Commission under contract LHCPhenoNet (PITN-GA-2010-264564), Madrid Regional Government (HEPHA- COS ESP-1473), Spanish Government (MICINN (FPA2010-17747)) and Spanish MINECO Centro de Excelencia Severo Ochoa Programme (SEV-2012-0249).}

\end{document}